\documentclass[12pt]{amsart}
\usepackage{graphicx} 
\usepackage{textgreek}
\usepackage{amsmath,amsaddr}
\usepackage[backend=bibtex,style=phys]{biblatex}
\usepackage{indentfirst,csquotes}
\begin{document}

\title{Plasmonic interferometer for nanoscale refractometry}
\author{Abbas Ghaffari and Robert Riehn\textsuperscript{\textdagger}}
\address{Department of Physics, NC State University, Raleigh, NC, USA}
\email{\textsuperscript{\textdagger}RRiehn@ncsu.edu}

\begin{abstract}We numerically study a double-slit interferometer that detects the modification of local phase and amplitude signal due to a refractive index contrast in a nanoplasmonic sensor after the plasmonic wave has been coupled to the farfield. Specifically, the sensor consists of two elongated zero-mode waveguides (ZMW) which function to launch surface plasmons as well as transmit light directly into the farfield. The surface plasmon waves are coupled to the farfield by surface gratings which cover one quarter of a bull-eye antenna, with different transmission directions. In the farfield, the coupled plasmon waves interfere with the directly transmitted reference wave, which reveals a combination of the phase and amplitude modification by the asymmetry of refractive index in the lumen of the ZMW. We report a drastic increase in sensitivity over interferometers that only utilize the directly transmitted fraction.
\end{abstract}
\maketitle
\section{Intro}
Sensitive, label-free detection of biological molecules enables the exploration of binding events in molecular biology without any modification of the target molecules, and thus free from possible artifacts that necessitate laborious repetition in proof-of-function controls. A contrast mechanism universal to all molecules is provided by the difference in refractive index between water and a biological molecule. However, because  single biological molecules in general are small when compared to the wavelength of light, either many molecules are required for successful sensing, or the effect of the biological molecule on a resonant optical system needs to be monitored, be that surface plasmons, and interferometric system with high finesse, or others.

In an earlier publication, we have introduced a sensor consisting of a zero-mode waveguide (ZMW) pair in a metal film that gives rise to a signal carried by  the direction of the transmitted radiation through the ZMW pair~\cite{ghaffari_nanophotonic_2023}. Specifically, we experimentally observed the detection of single quantum dots by using an optical system that was slightly defocused from a back-focal plane (Fourier) imaging mode, and explored the physics of the contrast mechanism. While the detector had the desired property of using the extremely small probe volume of a ZMW~\cite{levene_zero-mode_2003}, the sensitivity was not sufficiently high for detecting macromolecular assemblies or single molecule. We note that a very large fraction of light entering the ZMW-pair sensor was launched into plasmon polariton modes of the metal film that decay without being coupled to the farfield, and thus were not observable. Indeed, at short ZMW pair separations the performance deteriorated because of the close coupling of the ZMW through plasmonic modes. Here, we aim to recover the information that is carried through the interference of plasmon modes to significantly improve the sensitivity when compared to our initial sensor design.

The solution is based on the concept of bulls-eye antennas (BEA), which have been used extensively for coupling sub-wavelength apertures (another term for ZMW) or nano-emitters to the farfield with high efficiency~\cite{lezec_beaming_2002}. A BEA is a periodic series of concentric rings that surround a ZMW and act as an optical grating. If the radial propagation wavelength through the grating matches the grating period, all scattering events from the grating are in phase, and a directed beam of high intensity emerges~\cite{martin-moreno_theory_2003}. Under off-resonance conditions, a cone of light is emitted in an effect that is physically related to plasmonic phased arrays in beam steering applications~\cite{sattari_bright_2017,derose_electronically_2013}. By breaking the rotational symmetry of the BEA, complex polarization and spectral properties can be engineered~\cite{drezet_miniature_2008,kampouridou_complex_2021,butcher_all-dielectric_2022}. We follow this set of ideas by only using \textsuperscript{1}/\textsubscript{4} of a full BEA to selectively couple plasmons that travel along specific directions along the surface of the gold film.

In the following we will present a computational study of a nanoplasmonic device for sensing small particles based on their refractive index contrast with respect to water.  The devices use a ZMW pair to launch plasmons in the metal film and produce a plasmonic interference pattern of the surface wave. A pair of quarter bulls eye antennae (qBEA) couples the plasmon waves to the farfield, where they interfere with the directly transmitted light from the ZMW pair which forms a reference. We report a dramatic increase in device sensitivity over our prior publication, and estimate that the device would be able to detect a single 800\,kDa protein assembly. The present publication is the result of principles-guided design, and we anticipate that a subsequent algorithmic optimization will extend the sensitivity into the  single-protein range.

\section{Computational methods}
\begin{figure}
    \centering
    \includegraphics[width=1\linewidth]{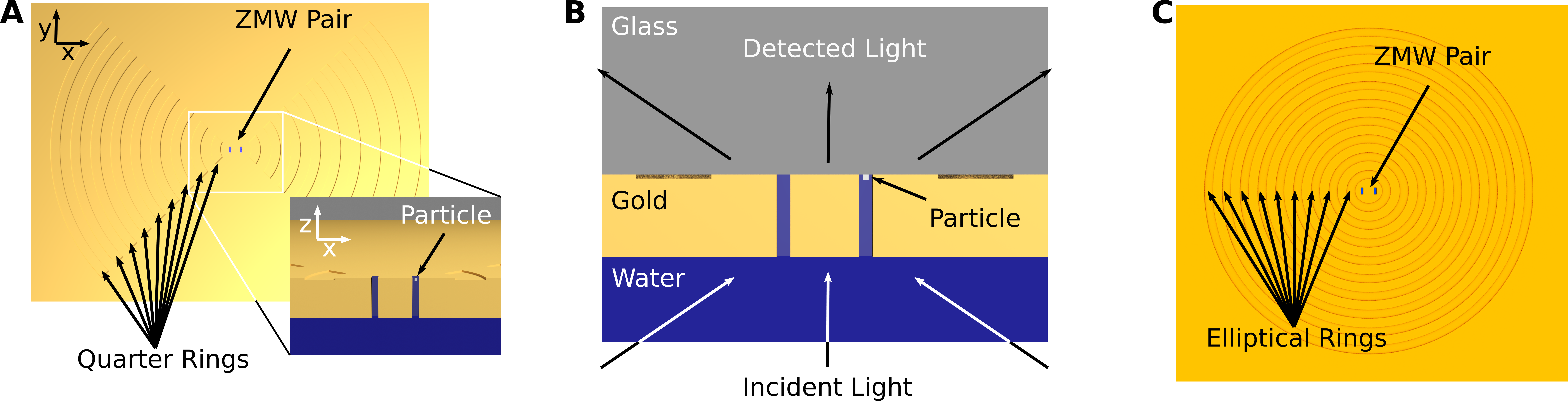}
    \caption{Schematic of the proposed device. \textbf{A} Glass-side view of detector with two ZMW, and one qBEA centered on each ZMW. The inset shows an oblique view of the center of the device. \textbf{B} Principle of operation. Light enters from the water side, traverses the gold film through the ZMW, and is coupled to the farfield. \textbf{C} Alternative design in which the ZMW pair is surrounded by ellipses.}
    \label{fig:Fig1}
\end{figure}

Fig.~\ref{fig:Fig1} illustrates the proposed device which is based on a pair of rectangular subwavelength apertures (Zero-mode waveguides, ZMW) in a 300\,nm-thick gold film. Unless specified differently, the apertures have a rectangular cross-section in the gold parallel to the substrate with edge lengths of 50\texttimes 150\,nm\textsuperscript{2}. The long edge of apertures is parallel to the $y$-axis, the displacement between the apertures is along the $x$-axis, and the origin is in the center between the ZMW and half-way through the gold film. The gold film is bounded by a silicon dioxide (glass) substrate on one side (drawn above the gold), and water on the other side (drawn below the gold). Water also fills the ZMW.

The interface between gold and glass is patterned with grating structures that protrude 30\,nm into the gold film. We illustrate quarter bulls eye antennas (qBEA) in Fig.~\ref{fig:Fig1}\textbf{A}, and elliptical bulls eye antennas in Fig.~\ref{fig:Fig1}\textbf{C}. The qBEA are formed by taking a quarter section of a bullseye antenna in which the grating has a rectangular cross-section (in a plane perpendicular to the substrate) with a 1:1 duty cycle of raised to depressed areas. The circles of one qBEA have one ZMW center at its center, and each ZMW has one associated qBEA. The distance from the center of the ZMW to the first edge of the qBEA is variable and an important design parameter. The elliptical bulls eye antennas are formed so that the innermost ring is an ellipse with mean major and minor radii so that the two ZMW lie in the focal points of the ellipse. All other rings are spaced so that the design grating periodicity is achieved along the two principal axes. In most cases nine periods of rings were used - a limitation that is imposed by the computation cell size. To test the ability of the device to detect objects trough refractive index contrast, we inserted a 20\texttimes 20\texttimes 20\,nm\textsuperscript{3} particle into one ZMW such that it was centered on the ZMW symmetry axis. Unless otherwise noted, the particle was located the at the gold-glass interface so that it is full enclosed by the water that fills the ZMW. The particle dimension was deliberately chosen such as to enable convenient meshing properties, without aiming at a specific physical system.

We used the finite-difference time-domain (FDTD) electromagnetic simulation approach as implemented in  Lumerical FDTD (Ansys Inc), a commercial software, to simulate the directional transmission pattern of the device. The refractive indices of glass and water were 1.43 and 1.33, respectively, and the refractive index of gold was given by a 6-parameter analytical fit to the experimental data published by Johnson and Christy~\cite{johnson_optical_1972}. The refractive index of the test particle was either 2.6 to represent a quantum dot particle~\cite{ninomiya_optical_1995}, or 1.43 to represent a protein cluster of about 800\,kDa molecular weight.

The simulation domain is a 6\texttimes 6\texttimes 3 \textmu m\textsuperscript{3} box surrounded by perfectly matched layers. An adaptive mesh with mesh size of about 5\,nm in the ZMW region spans the simulation domain. Satisfactory convergence under change of meshing parameters was observed. A Gaussian source at normal incidence polarized in the $x$-direction with a waist size of 800\,nm at wavelengths covering 650\,nm to 850\,nm illuminates the sample from the water side, where the beam waist was located on the water-gold interface. During each 200-fs simulation, fields were collected on monitor planes located in the glass, 150\,nm and 450\,nm from the glass-gold interface. The farfield transmission patterns were obtained from complex-valued nearfield-to-farfield transformation at the plane further from the gold film, while the plane closer to the gold film was used to monitor the propagation of plasmons. The directional dependence of transmitted light was plotted as the cosine of the angle of the unit vector of propagation $\hat k$ with the Cartesian axes, i.e. $u_x=\hat x\cdot \hat k$ and $u_y=\hat y\cdot \hat k$. The relatively large separation of the monitor plane from the gold film leads to loss of information about high exit angles. However, since all information essential to the functioning of the sensor as an interferometer is within a 30\textdegree cone from the normal, we consider that a fair trade-off if it reduces the susceptibility to oscillations that could result from a nearfield-to-farfield transformation that mainly probes non-radiating surface-bound waves. 

\section{Results and discussion}
\subsection{Proof of Principle}
\begin{figure}
    \centering
    \includegraphics[width=\linewidth]{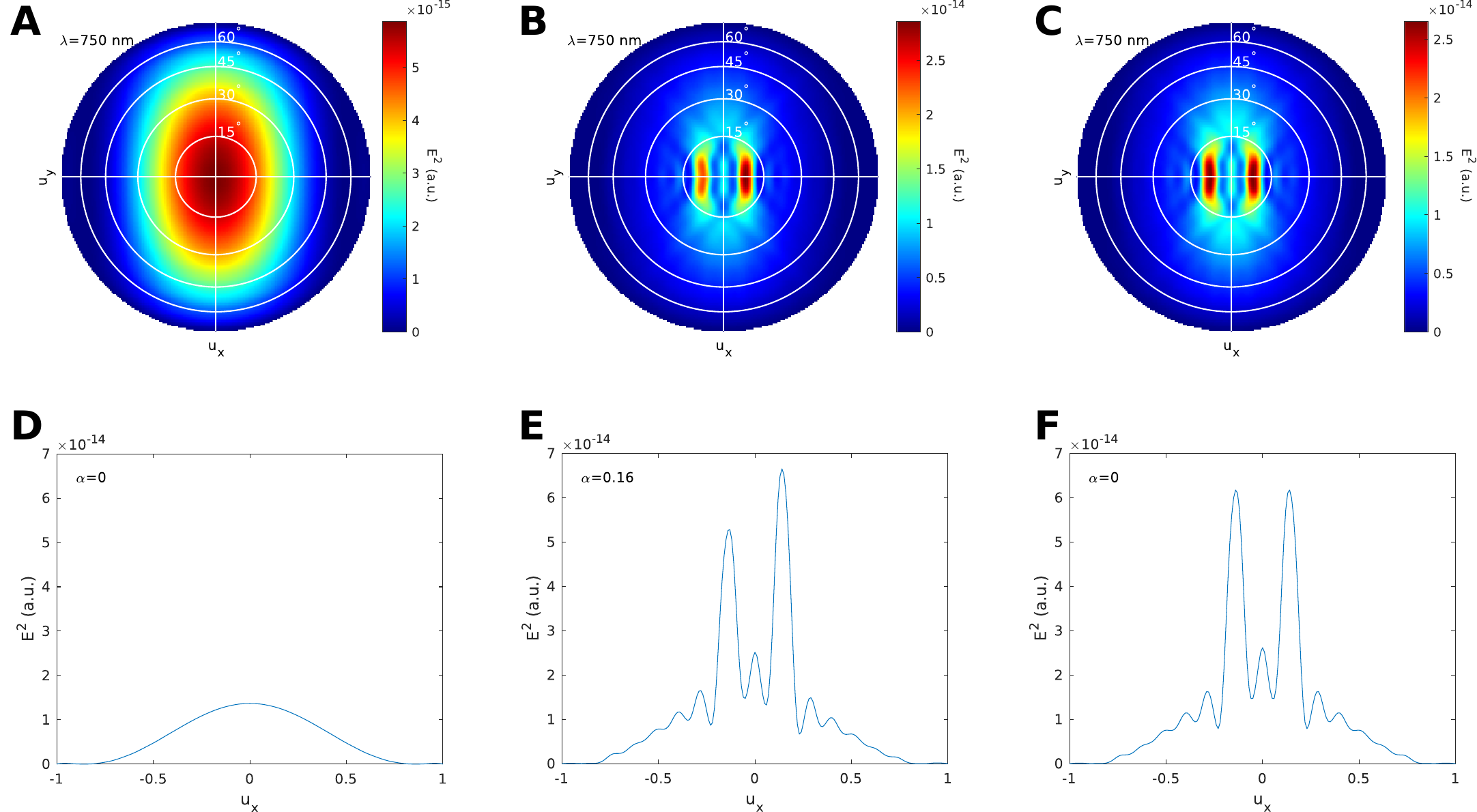}
    \caption{Illustration of interferometric detection by devices with 300\,nm Au thickness, 50$\times$150\,nm\textsuperscript{2} ZMW size, and 300\,nm center-to-center spacing at 750\,nm wavelength. The plasmonic antennae were two qBEA with 9 rings that are centered each on one ZMW, with a 550\,nm period, and a distance of 412.5\,nm form the center of the nearest ZMW to the first edge of the antenna. One ZMW contains a (20\,nm)\textsuperscript{3} particle with $n=2.6$. \textbf{A-C} Electric field distribution versus direction in the $u_x$/$u_y$ plane. \textbf{D-F} Quantification of asymmetry at $u_y=0$. \textbf{A, D} Without plasmonic antenna. \textbf{B, E} With plasmonic antenna. \textbf{C, F} Control with plasmonic antenna but no particle.  }
    \label{fig:Sales}
\end{figure}

We establish the basic functionality of the detector in Fig.~\ref{fig:Sales}. Here we compare the transmission direction pattern of (i) a detector with a ZMW pair where one ZMW contains QDot-like particle as in our prior publication (Fig.~\ref{fig:Sales}\textbf{A}), (ii) a detector with a ZMW pair and two qBEA with the particle in one ZMW (Fig.~\ref{fig:Sales}\textbf{B}), and a detector with a ZMW pair and two qBEA without a particle (Fig.~\ref{fig:Sales}\textbf{C}). We observe a broad, featureless peak without the qBEA (Fig.~\ref{fig:Sales}\textbf{A}), while two distinct peaks along the $u_y=0$ axis appear with the qBEA (Fig.~\ref{fig:Sales}\textbf{B,C}). In Fig.~\ref{fig:Sales}\textbf{D-E} we present sections through Fig.~\ref{fig:Sales}\textbf{A-C} at  $u_y=0$. By comparing Fig.~\ref{fig:Sales}\textbf{E} and \textbf{F}, we clearly see that the particle in Fig.~\ref{fig:Sales}\textbf{B,E} gives rise to a clear asymmetry between the two peaks, while the broad background is disturbed by ripples. The minute asymmetry without the qBEA (Fig.~\ref{fig:Sales}\textbf{A,D}) can not be seen on visual inspection. 

The remainder of the results section will elucidate the mechanism that gives rise to this signal. However, we can already make a few key observations. First, the lack of large-scale deflection or asymmetry without qBEA in Fig.~\ref{fig:Sales}\textbf{A,D} is consistent with our prior reports. There, we defocused the detection optics to obtain a phase-sensitive ripple pattern.  The second is that the patterns with qBEA are the interference pattern of the broad feature of light directly transmitted by the ZMW pair, and a two-peak pattern due to a surface-bound plasmon polariton wave that is launched into the farfield by the BEA. Importantly, the interference is evidently coherent and phase sensitive as we see one of the peaks in Fig.~\ref{fig:Sales}\textbf{E} rise and one fall when comparing to Fig.~\ref{fig:Sales}\textbf{F}.

\subsection{Design of detector}
An optimal design would be based on numerical optimization of all tunable parameters of the detector geometry, using methods that have been developed in similar contexts~\cite{wang_intelligent_2021}. However, those methods are most powerful if they are guided by a physical intuition and a productive initial configuration so that a search does not become trapped in local optimal points far from the global optimum. Thus, we considered the the detector as a collection of discrete units that would be combined using established principles. Our results are thus not meant to show the ultimate limits of the technology, but rather the basic capability.

\subsubsection{Single Holes}
\begin{figure}
    \centering
    \includegraphics[width=0.8\linewidth]{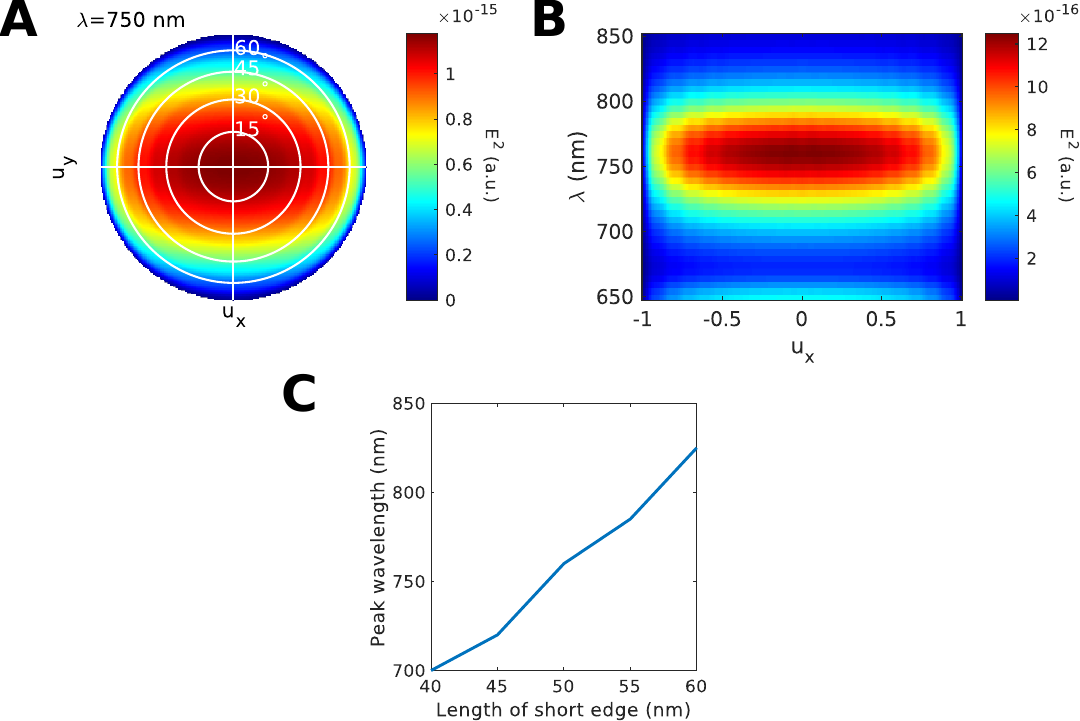}
    \caption{Resonances in single-ZMW detector without bulls eye antennae. \hbox{\textbf{A} Transmission} of single 50$\times$150\,nm\textsuperscript{2} ZWM at 750\,nm wavelength.  \textbf{B} Profile of the transmission of the same ZMW as in \textbf{A} in the $u_x$ \textit{vs.}\ $\lambda$ plane at $u_y=0$. \textbf{C}\ Wavelength of highest transmission versus edge length for ZMW with an $l_y:l_x$ aspect ratio of $1:3$.}
    \label{fig:SingleNoRing}
\end{figure}

In our prior publication we used circular ZMW for ease of fabrication. However, established reports show that subwavelength-width slits excited by light polarized perpendicular to their long axis have a resonance  reminiscent of a Fabry-Perot interferometer~\cite{sain_plasmonic_2018}. To utilize that idea, we fixed the gold thickness at 300\,nm, fixed aspect ratio at 1:3 ($x$-extent:$y$-extent), and varied edge length. We find a size-dependent resonance (Fig.~\ref{fig:SingleNoRing}) whose resonance wavelength is taken as the wavelength of the maximum farfield-transmitted electric field amplitude. For the purpose of this paper, we target a detector operating using 750\,nm-wavelength illumination from a diode laser, and thus fixed the edge length in the $y$-direction at 50\,nm and the $y$-direction at 150\,nm. Higher aspect ratios and sharper resonance could be achieved, but would lower the likelihood that a particle binds at the center of the gold/glass interface within a ZMW.

\subsubsection{Hole Spacing}
\begin{figure} 
    \begin{center}
        \includegraphics[width=\linewidth]{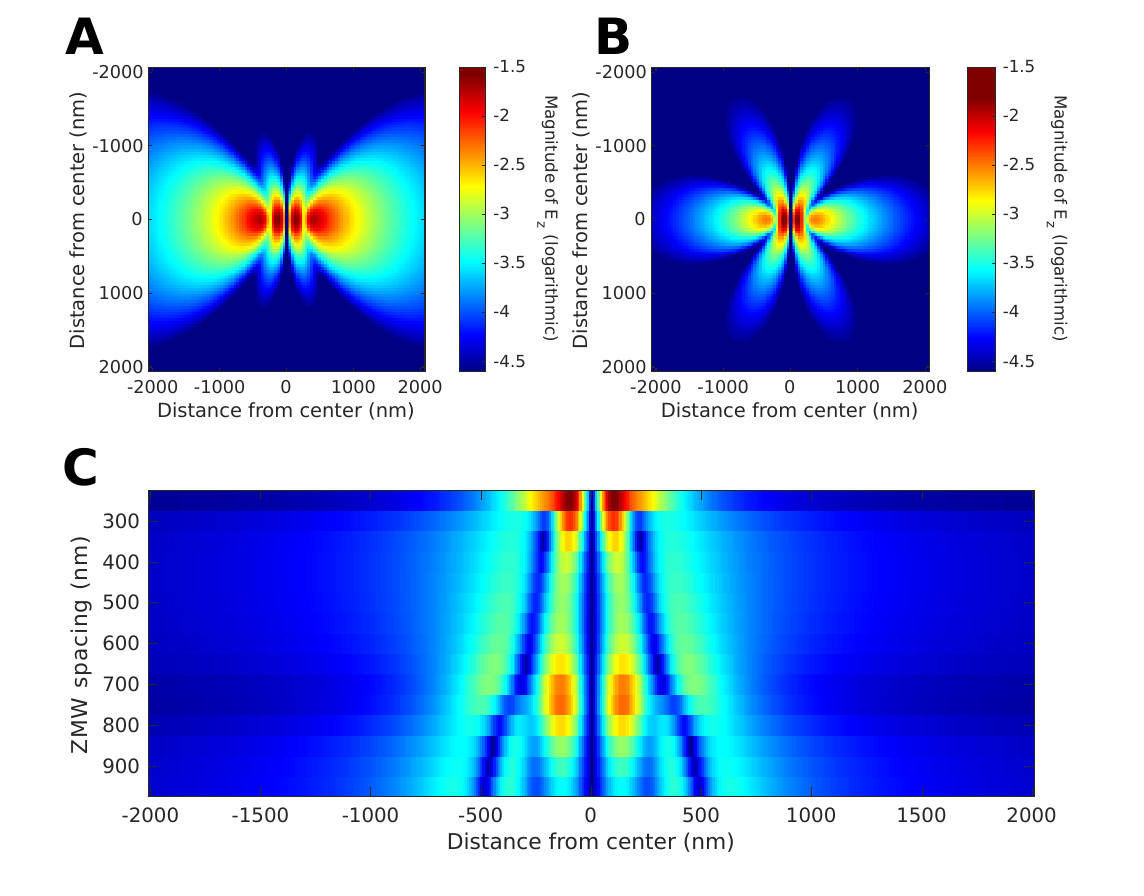}
    \end{center}
    \caption{Resonances in twin-ZMW detectors without BEA with ZMW geometry fixed at 50$\times$150\,nm\textsuperscript{2} and 300\,nm Au thickness. All analysis was performed at a wavelength of 750\,nm by observing the magnitude of the $z$-component of the transmitted electric field in a plane 150\,nm into the glass substrate. \textbf{A} Logarithm of the magnitude for a center to center separation of 500\,nm. \textbf{B} Logarithm of the magnitude for a center to center separation of 300\,nm. \textbf{C} Section through panels such as \textbf{A} and \textbf{B} for practical hole separations at $y=0$\,nm. The trace was normalized for each separation value to allow easy comparison.}
    \label{fig:TwinNoRing}
\end{figure}

Plasmon polariton waves travel along the gold surface, and are subject to interference effects there. Our detector has two ZMW and we need to consider the impact of their spacing on the magnitude and the distribution of the launched plasmons (Fig.~\ref{fig:TwinNoRing}). As a proxy for the strength of the plasmon we take the $z$-component of the electric field measured in the glass $150$\,nm from the gold-glass interface. Due to antisymmetry along the $x$-axis under $x$-polarized illumination, the $z$-component must vanish at $x=0$\,nm. Similarly, it has a local minimum above each ZMW.  

For a ZWM spacing of $S=500$\,nm, two prominent plasmon jets along the $x-axis$ emerge to the left and the right of the pair, and four smaller lobes extend above and below the $x-axis$. (Fig.~\ref{fig:TwinNoRing}\textbf{A}). We observe a similar intensity between the ZMW and outside of the ZMW pair when moving along the x-axis.   For a ZMW spacing of $S=300$\,nm, the prominence of the jets along the x-axis is greatly diminished, and the local maxima between the ZMW become dominant in a resonance while the diagonal lobes above the $x$-axis extend further (Fig.~\ref{fig:TwinNoRing}\textbf{B}). Systematic investigation of the effect shows that the it can be understood in terms of interference of the plasmon waves from the two ZMW, which we will illustrate for a vacuum wavelength of \textlambda=750\,nm (Fig.~\ref{fig:TwinNoRing}\textbf{C}). At that vacuum wavelength, the plasmon polariton wavelength is \textlambda\textsubscript{spp}=490\,nm using our parametrization of the FDTD simulation. At \textit{S} = \textlambda\textsubscript{spp} we anticipate destructive interference of plasmons in the region between the ZMW, or an anti-resonance. At \textit{S} = \textsuperscript{3}/\textsubscript{2} \textlambda\textsubscript{spp} and \textit{S} = \textsuperscript{1}/\textsubscript{2} \textlambda\textsubscript{spp} we anticipate constructive interference in the region between the ZMW, or resonance. The interference phase on the outside of the region between the ZMW is phase-offset by about $\pi/2$ relative to the one between the ZMW, and so we anticipate a minimum in the amplitude of the  plasmon wave launched along the $x$-direction exactly at the resonance condition. Crucially, the plasmons under the resonance condition are sensitive to both holes, even if they appear to emerge from left or right hole.

\subsubsection{Quarter Bulls-Eye Antennas}
\begin{figure}
    \centering
    \includegraphics[width=\linewidth]{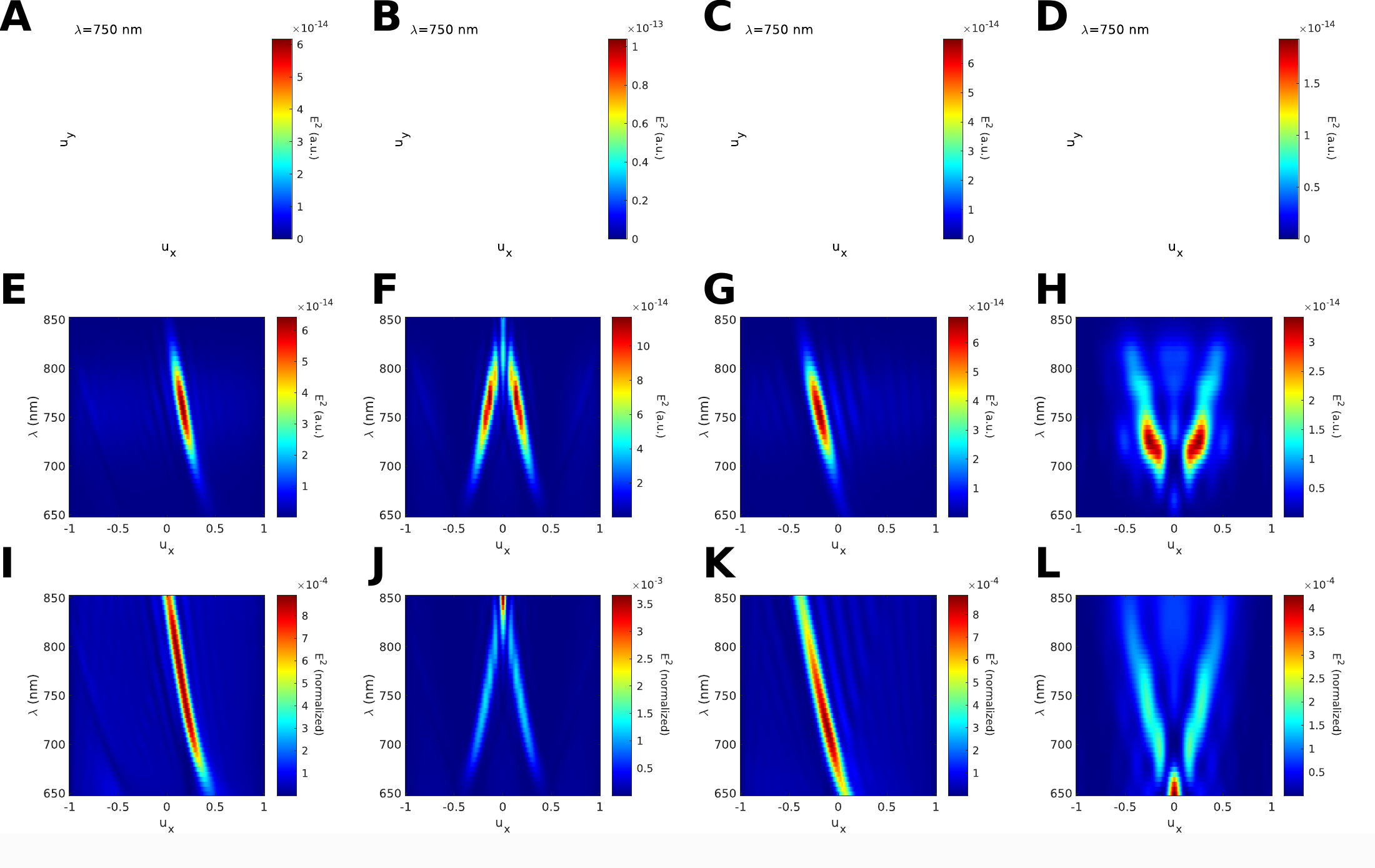}
    \caption{Design of qBEAs and pairs of qBEAs with a grating modulation depth of 30\,nm.  \textbf{A-D} Transmission patterns at 750\,nm. \textbf{E-H} Section through emission pattern \textit{vs.}\ wavelength at $u_y$=0. \textbf{I-L} Section through emission pattern \textit{vs.}\ wavelength at $u_y$=0 after intensity normalization at each wavelength. \textbf{A,E,I} Single ZMW with one qBEA with 550\,nm grating period. \textbf{B,F,J} Two ZMWs spaced by 550\,nm with one qBEA each with 550\,nm grating period.  \textbf{C,G,K} Single ZMW with one qBEA with 400\,nm grating period. \textbf{D,H,L} Two ZMWs spaced by 300\,nm with one qBEA each with 400\,nm grating period.}
    \label{fig:BEA}
\end{figure}

Single source BEAs with full circles have been widely reported in the literature. Segmented BEA supporting multiple resonance modes or non-isotropic BEAs to manipulate polarization have also been reported~\cite{butcher_all-dielectric_2022,drezet_miniature_2008}. The  plasmon jets from twin ZMW detector that were observed in Fig.~\ref{fig:TwinNoRing} imply that we can employ sectors of BEAs without a loss of coupling of plasmons to the farfield. We thus selected qBEAs for the majority of the remaining studied. Commonly BEAs are used in a beaming configuration where a single beam emerges perpendicular to the substrate. However, they can also be used for plasmonic beam steering if used off-resonance (circles leading to cones~\cite{pu_near-field_2015}), and that is the mode that we will employ here. In an off-resonance configuration, we can use either a wavelength below the beaming condition, or above the beaming condition, both of which are shown in Fig.~\ref{fig:BEA}.

In particular, we demonstrate that a qBEA with a single rectangular ZMW leads to a single well-defined dot in the $u_x$/$u_y$ plane (Fig.~\ref{fig:BEA}\textbf{A,C}). For a grating period of 400\,nm and a modulation depth of 30\,nm the beaming condition is given by $\lambda\simeq$650\,nm (Fig.~\ref{fig:BEA}\textbf{E,I}), and for a grating period of 550\,nm the beaming condition is $\lambda\simeq$850\,nm (Fig.~\ref{fig:BEA}\textbf{G,K}). We interpolate that a grating period of $\simeq$475\,nm would give beaming at our detector design wavelength $\lambda\simeq$750\,nm, which naturally must coincide with the \textlambda\textsubscript{spp} that was inferred from Fig.~\ref{fig:TwinNoRing}.

Using symmetric detectors with two ZMWs and two qBEAs, each focused on one ZMW, approximately leads to a symmetric pattern containing two mirror images of the single-ZMW device (Fig.~\ref{fig:BEA}\textbf{B,D}), with beaming conditions unchanged from the single-ZMW/qBEA cases (Fig.~\ref{fig:BEA}\textbf{F,H,J,L}). However, they also show the emergence of the interference of light directly coupled from the ZMWs to the farfield with the light that is first coupled into the plasmon and then scattered into the farfield by the qBEA grating.

\subsubsection{Twin-ZMW devices with qBEAs as detectors}
In Fig.~\ref{fig:Sales} we had demonstrated that a twin-ZMW detector with qBEAs is sensitive to a modulation of the refractive index in one of the two apertures. We are now in a position to interpret how this contrast arises. We have now shown that the pattern is the result of the interference of the directly transmitted light with light scattered out of the plasmon wave emanating from the ZMW pair either to the left or the right. We can make a number of conscious design decisions at this point. Below, we explore the phase of the ZMW-coupled light as well as the impact of the ZMW spacing.

\begin{figure}
    \begin{center}
        \includegraphics[width=0.7\linewidth]{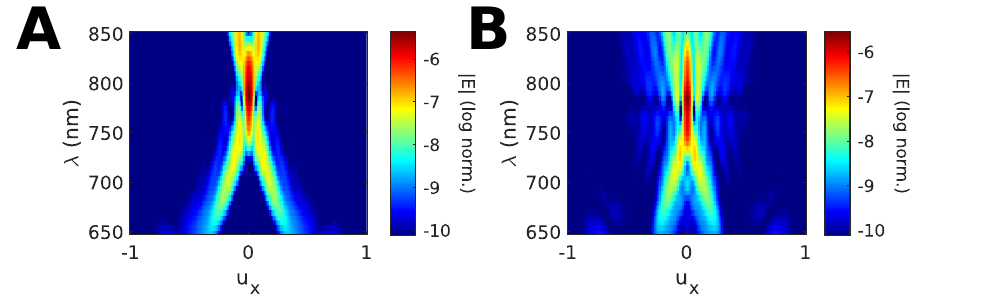}
    \end{center}
    \caption{Logarithm of farfield electric field intensity at $u_y=0$, with the intensity in each line normalized for visibility, for a single aperture and a grating period of 500\,nm. In \textbf{A} the inner ring edge is  located at 375\,nm while in \textbf{B} it is located at 625\,nm.\label{fig:SwitchPhase}}
\end{figure}

While the scattering angle of the qBEA-scattered light is given by the period of the grating (Fig.~\ref{fig:BEA}), the distance of the first ring of the grating to its focal ZMW sets the phase of the scattered wave. 
A single FDTD sweep using a single ZMW with a full BEA with period of 500\,nm contains wavelengths both below and above the beaming condition (Fig.~\ref{fig:SwitchPhase}). The clearest manifestation of the phase effect is at the shorter-than-beaming wavelengths, where at a location of the first ring from 375\,nm (Fig.~\ref{fig:SwitchPhase}\textbf{A}) the bright cone is delineated by a dark region at smaller $u_y$. At a location of the first ring 625\,nm (Fig.~\ref{fig:SwitchPhase}\textbf{B}), the bright cone occupies the previously bright region while the previously dark region now contains the bright cone. We note that there certainly is a plasmon wave that is reflected at the outer edge of the BEA and that then interferes with the light scattered from the primary plasmon at the other side of the ZMW, and so the discussion above should be seen as a simplified conceptualization. 

\begin{figure}
    \centering
    \includegraphics[width=0.7\linewidth]{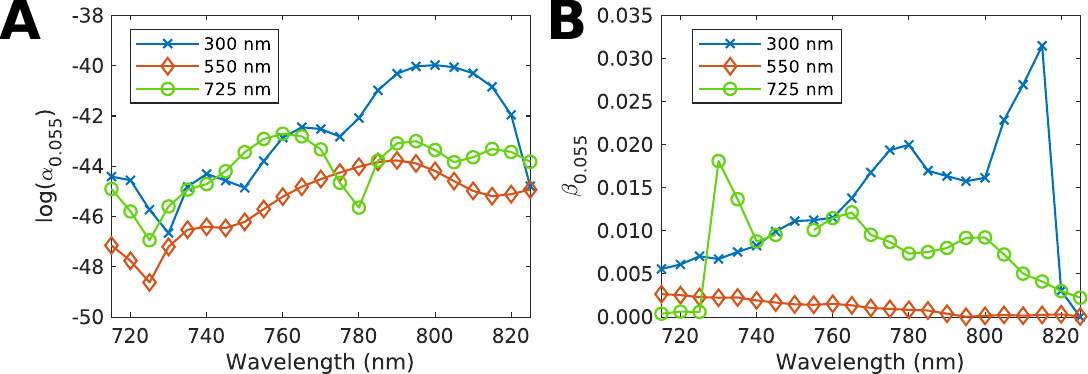}
    \caption{Quantitative asymmetry measures for select ZMW separations. qBEAs have a period of 550\,nm with a separation from ZMW center to first BEA edge of 412.5\,nm. Panels show logarithm of $\alpha_{0.055}$ measure (\textbf{A}), and logarithm of $\beta_{0.055}$ measure (\textbf{B}). Curves are shown for 300\,nm separation (blue \texttimes), 550\,nm separation (red $\diamond$), and 725\,nm (green $\circ$).}
    \label{fig:SeparationBowtie}
\end{figure}

We now discuss the impact of the hole spacing. The  ZMW pair separations of exceptional interest are those that were identified in  Fig.~\ref{fig:TwinNoRing} as having a resonance of the $E_z$-field versus those that were non-resonant. At the resonance, we observed destructive interference of the plasmons within the jets, and thus the potential for high sensitivity. To quantify such a statement, we have to introduce a numerical measure of the asymmetry of the distribution of light observed in the farfield. At first, we choose an computationally simple measure as
\begin{equation}
    \alpha_w=\frac{\int\limits_{u_x=-1}^{1}\mathrm d u_x\int\limits_{u_y=-w/2}^{w/2} d u_y \left(I(u_x,u_y)-I(-u_x,u_y)\right)^2}{\int\limits_{u_x=-1}^{1}\mathrm d u_x\int\limits_{u_y=-w/2}^{w/2} d u_y \left(I(u_x,u_y)+I(-u_x,u_y)\right)^2}\,\,,
\end{equation}
assuming data is measured as $I_{x,y}$ where the array has $n\times n$ points, and the width of the analyzed stripe around $u_y=0$ is $w$.  In Fig.~\ref{fig:SeparationBowtie}\textbf{A} we plot $\alpha_{0.025}$, and find that the small resonant hole separation (300\,nm) yields by far the greatest maximum asymmetry, followed by the large resonant hole separation (725\,nm), while the off-resonance hole separation (550\,nm) has the smallest asymmetry. Note that these curves are not monotonous as a function of wavelength, but rather show oscillations.

\begin{figure}
    \centering
    \includegraphics[width=0.5\linewidth]{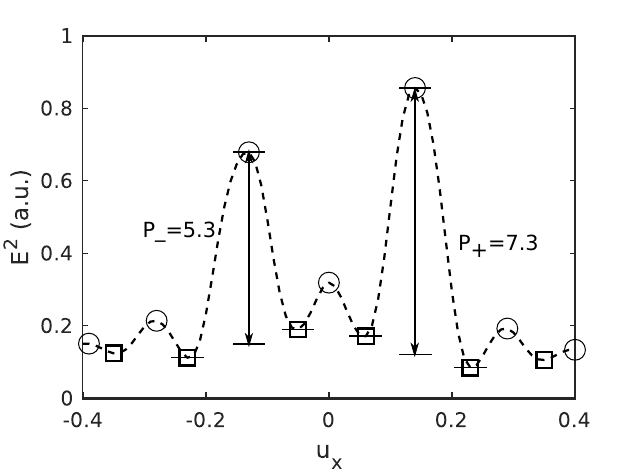}
    \caption{Finding the $\beta_{0.055}$ measure for the data in Fig.~\ref{fig:Sales}\textbf{B,E}. The dashed line is the integral over $u_y$ from $-w/2$ to $w/2$, circles show maxima, and squares show minima. The prominence of a maximum is the average difference to its two proximal minima.}
    \label{fig:annotate}
\end{figure}

In a practical setting, the $\alpha$ measure requires the absolute knowledge of the symmetry line of the empty detector, which may be difficult to obtain in many camera-based experimental setups. A computationally cheap and model-free measure  to determine the asymmetry is the prominence of the tallest peak to the left and the right of a possible central peak ($P_-$ and $P_+$, respectively), after averaging over $u_y=-w/2\hdots+w/2$. If the prominence $P_i$ of the $i$\textsuperscript{th} peak is the average difference between the peak amplitude and the amplitude of the neighboring minima, we define
\begin{equation}
    \beta_w=
    \frac{P_+-P_-}{P_++P_-}\,\,.
\end{equation}
That process is illustrated in Fig.~\ref{fig:annotate}. Clearly, if more than three prominent peaks are present, or the pattern becomes chaotic, the $\beta$ measure can become unstable in itself. Applying the $\beta$ measure to the question of the ZMW separation leading to maximal sensitivity, we similarly find that the 300\,nm separation is the best starting point for an optimization. Note that a different phase of the rings may recover some sensitivity even for 550\,nm separation, but such an exhaustive search was out of our scope here.

\begin{figure}
    \centering
    \includegraphics[width=0.8\linewidth]{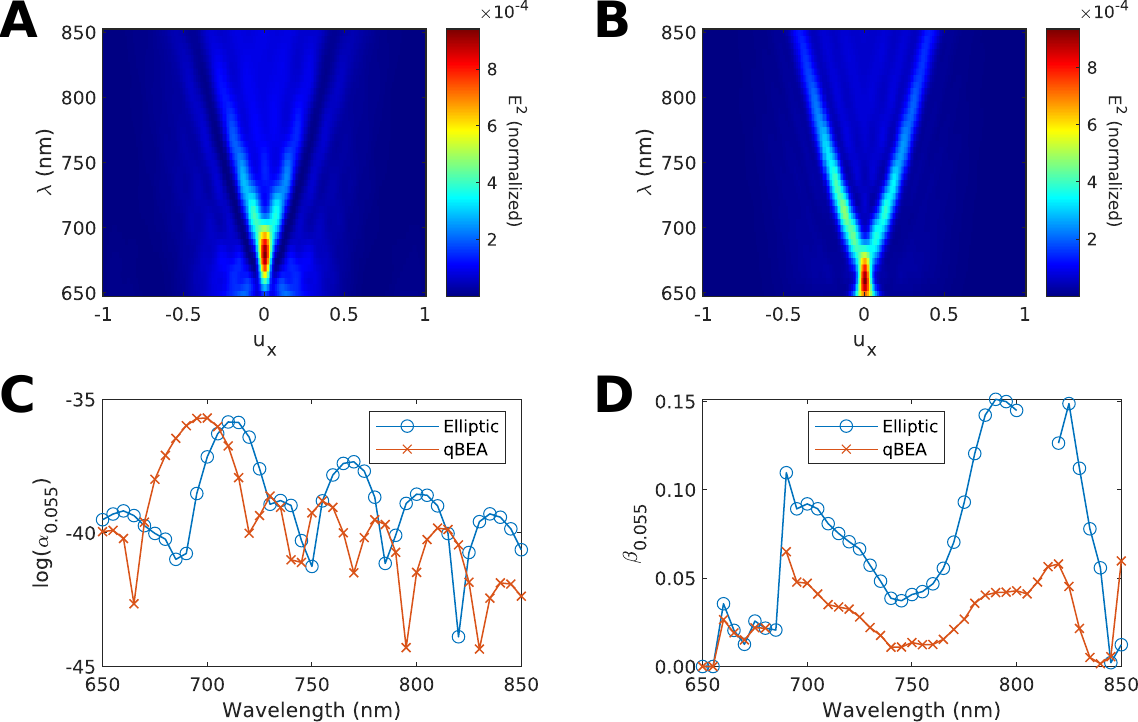}
    \caption{Comparison of antenna designs with 300\,nm ZMW spacing, 400\,nm grating period, and a $20\times20\times20$\,nm\textsuperscript{3} particle with $n=2.6$ in one ZMW. \textbf{A} Elliptic antenna (R500). \textbf{B} qBEA antennas (R500). \textbf{C} $\alpha_{0.055}$ measure of asymmetry and \textbf{D} $\beta_{0.055}$ measure of asymmetry. Values of failure to identify peaks omitted. Blue $\circ$ indicate the data from panel \textbf{A} and red \texttimes indicate the data from panel \textbf{B}. }
    \label{fig:BowtieEllipse}
\end{figure}

Up to this point, we followed the design idea based on two qBEA, one focused on each hole, to avoid the problem of full circular BEA that the two apertures in a pair cannot each be at the focal point of the same set of circles. However, this is not the only possible design choice. For instance, we have investigated elliptical antennas that have the advantage of fully encircling the pair. In Fig.~\ref{fig:BowtieEllipse} we compare elliptical antenna and the twin-qBEA configuration and find that at wavelengths just longer than the beaming condition (here $\simeq$675\,nm), both perform similarly. However, at large offsets from the beaming condition the $\beta$ measure failed more frequently for the elliptic antenna design. Even though the elliptic design is not inherently inferior, we selected the qBEA design because the results can be interpreted more readily. Note also that our choice of using a detector a wavelengths shorter than the beaming condition in Fig.~\ref{fig:Sales} was arbitrary at first.

\begin{figure}
    \centering
    \includegraphics[width=0.8\linewidth]{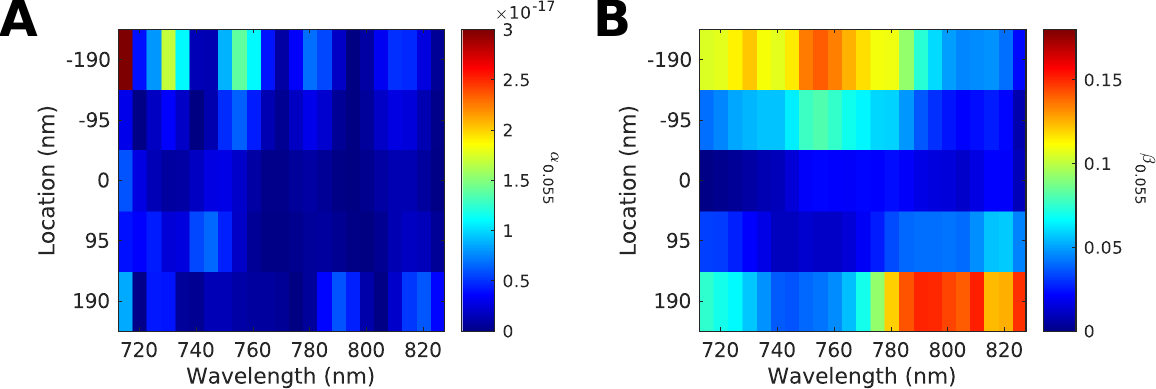}
    \caption{Sensitivity of devices as function of location of a $20\times20\times20$\,nm\textsuperscript{3} particle with $n=2.6$ in one ZMW while varying location of the particle along axis of ZMW. Device has with 300\,nm ZMW separation, an elliptic antenna with 400\,nm period, and a distance from ZMW center to first antenna edge of 250\,nm. We present both $\alpha_{0.055}$ (\textbf{A}) and $\beta_{0.055}$ (\textbf{B}). The location axis shows the center of mass of the particle with respect to the middle of the ZMW, so that negative numbers are closer to the gold/glass interface while positive numbers are towards the gold/water interface.}
    \label{fig:ZScan}
\end{figure}

For a future application, we envision that a biological sample will be tethered at the ``bottom'' of one of the ZWM. In that context, it is important to know whether free particles could also lead to a signal. We explored this aspect by moving a particle along the axis of a ZMW (Fig.~\ref{fig:ZScan}) when the ZWM dimensions are resonant as found in Fig.~\ref{fig:SingleNoRing}. We find that the $\alpha$ measure indicates that the highest sensitivity is obtained at the gold/glass interface, no sensitivity is found in the center, and some sensitivity is recovered at the gold/water interface. The $\beta$ measure paints a similar picture, but has a higher sensitivity at the gold/water interface. Interestingly, the wavelengths at which the maximum sensitivities are obtained are different for the two interfaces. Both the $\alpha$ and $\beta$ measures give similar trends, but different magnitudes and details. Using the $\alpha$ measure, shorter wavelengths lead to a dramatically increased sensitivity at the gold/glass interface, while the sensitivity at the gold/water interface suffers a broad window of insensitivity between 730\,nm and 780\,nm. Using the $\beta$ measure, a particle at the gold/glass interface is most efficiently detected at 755\,nm, while a particle at the gold/water interface is most efficiently detected at 790\,nm. At 775\,nm both locations have a similar sensitivity.  We conclude that both interfaces are sensitive to detection, the lumen is insensitive, and the choice of wavelength for the specific detector can lead to a dramatic selectivity of the interface that is being probed.

\subsection{Limits of sensitivity}
We will now explore the limits of sensitivity that are attainable with a manual exploration, that is without simultaneous numerical optimization of multiple parameters. To do so, we replace the particle used up to this point with one that has the same size ($20\times20\times20$\,nm\textsuperscript{3}), but a refractive index of 1.43. Using this choice, we avoid problems in meshing objects that are too small, and we introduce a refractive index change equivalent to about a single 1\,MDa protein complex~\cite{khago_protein_2018}. Because a significant fraction of light in this case is scattered at $\left|u_y\right|>0$, we chose quantification of the asymmetry with $w=0.104$. 

\begin{figure}
    \centering
    \includegraphics[width=0.8\linewidth]{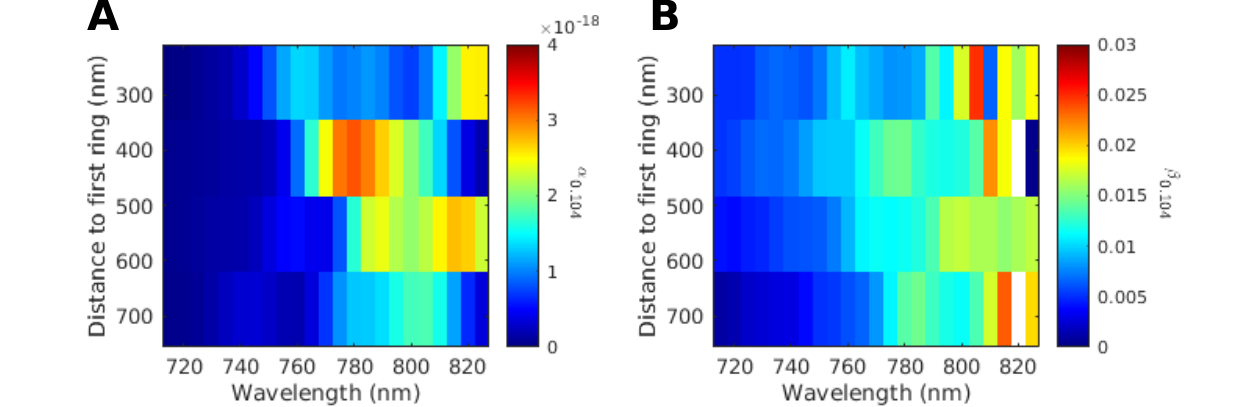}
    \caption{Asymmetry measures for devices with 300\,nm hole separation and 550\,nm grating period as function of distance from center of ZMW to first edge of qBEA. One ZMW contained a $20\times20\times20$\,nm\textsuperscript{3} particle of $n=1.43$ at the gold/glass interface. \textbf{A} The $\alpha$ measure shows maximal response at a distance of 412.5\,nm. \textbf{B} The $\beta$ measure contains some quantification failures close to $\lambda=800$\,nm.}
    \label{fig:SensPhase}
\end{figure}

The starting point of the optimization is the 300 nm ZMW separation detector that was the basis of Fig.~\ref{fig:Sales}. We first test whether the distance of the first edge of the qBEA to the center of the nearest ZMW is optimal (Fig.~\ref{fig:SensPhase}). We find a clear optimum of the $\alpha$ measure if the first edge is about 3/4 of a grating period located from the center of the ZMW. An oscillation pattern is observed at the other distances detested that is consistent with the interpretation that the asymmetry emerges from the interference of the directly transmitted light and the scattered surface plasmon. 

\begin{figure}
    \centering
    \includegraphics[width=0.8\linewidth]{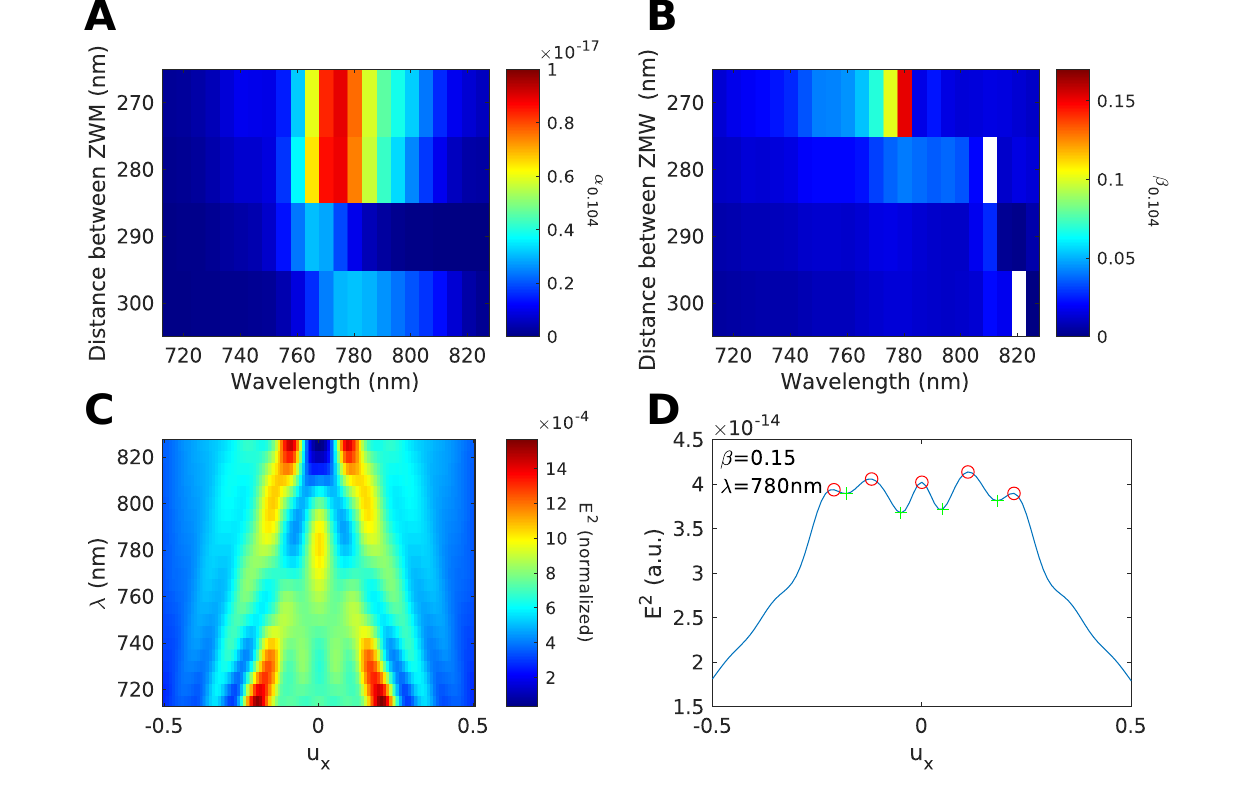}
    \caption{Optimizing the center-to-center ZMW separation for maximum sensitivity with qBEA period of 550\,nm and distance from ZMW center to first qBEA edge of 412.5\,nm. One ZMW contained a $20\times20\times20$\,nm\textsuperscript{3} particle of $n=1.43$ at the gold/glass interface. \textbf{A} $\alpha$-measure shows high sensitivity for both 270\,nm and 280\,nm ZMW separation. \textbf{B} $\beta$-measure shows highest sensitivity for 270\,nm separation. \textbf{C} Zoom of line-normalized transmitted intensity at 270\,nm separation after averaging over the $u_y$ range underlying the $\alpha_{0.104}$ and $\beta_{0.104}$ calculations. \textbf{D} Identification of peaks for $\beta_{0.104}$ at $\lambda=780$\,nm and 270\,nm separation.}
    \label{fig:SensDist}
\end{figure}

Following that idea, we anticipate that the highest sensitivity is obtained if the intensity of both contributions is comparable. In Fig.~\ref{fig:Sales}, the contribution from the scattered plasmon was clearly dominant. In order to increase the relative importance of the directly transmitted light we reduce the distance between the ZMW (Fig.~\ref{fig:SensDist}). We find that the asymmetry measure becomes increasingly sensitive as the distance between the ZMW is reduced, with a dramatic increase of $\alpha_{0.104}$ for separations of both 270\,nm and 280\,nm at wavelengths around 770\,nm. We find that $\beta_{0.104}$ is the highest 270\,nm separation, where we obtain $\beta_{0.104}=0.15$ at a wavelength of $780$\,nm.  That is comfortably larger than the 0.1 of a related asymmetry measure that we had set as a detection threshold in an earlier publication~\cite{ghaffari_nanophotonic_2023}. However, we can clearly identify that a further development of a more robust asymmetry measure is necessary.

\section{Conclusion and Outlook}
In this publication we have introduced a modified nanophotonic interferometer that utilizes both light directly transmitted through a pair of ZMW as well as light that is scattered out of a plasmon wave by qBEA. In our computational study, we found that refractive an refractive index contrast equivalent to about 10 large proteins in one of the apertures can be detected. To so, we fixed a desired working wavelength (750\,nm) as well as the metal film thickness, and then adjusted the aperture size, ZMW spacing, qBEA period, and phase in a sequential (i.e.\ non-iterative) fashion to yield the most sensitive configuration.

Throughout this manuscript, we have emphasized that the present design is a rationally designed sensor before the necessary computational optimization that must follow as a next step. We believe that such an optimization can lead to a detection sensitivity equivalent to single large proteins. In particular, one needs to optimize the Au film thickness, introduce a non-rectangular aperture geometry, possibly use non-circular qBEA, as well as vary the qBEA modulation depth, duty cycle, and phase. We also note that our asymmetry quantification was not sophisticated, and therefore an efficient design has to be developed in lock-step with a suitable asymmetry quantification that is both sensitive and robust.

Finally the implementation of the concept developed here is in principle achievable using the methods of our earlier paper, with the only technological difference being the grating that needs to be established using aligned multi-exposure nanolithography of the qBEA at the interface of SiO\textsubscript{2} and Au. The optical detection, however, is identical as Fourier-imaging of a transmission confocal microscope. The ultimate detection limit is likely not going to be limited by the design of the  detector, but rather fabrication variation that leads to asymmetries that could far exceed those induced by macromolecular analytes. Future usability is further impacted by the ability to localize a particle at a specific point in the H\textsubscript{2}O-SiO\textsubscript{2} interface.

\section*{Acknowledgements}
We acknowledge support from the National Institutes of Health (GM126887) and US Air Force Office of Scientific Research (AFOSR, FA9550-23-1-0311).

\section*{Conflicts}
The authors declare no financial conflicts.

\section*{Author contributions}
AG and RR conceived and conducted the research and prepared the manuscript.

\section*{Data availability}
Data underlying published figures is available upon reasonable request.

\printbibliography
\end{document}